%Paper: gr-qc/9412043
%From: imtjj51 <imtjj51@cc.csic.es>
%Date: Thu, 15 Dec 1994 18:17:34 UTC+0200

% file translat.tex, 27-06-1994, Tresguerres
\magnification=\magstep1
\hsize=13cm
\vsize=20cm
\overfullrule 0pt
\baselineskip=13pt plus1pt minus1pt
%\baselineskip=2\baselineskip
\lineskip=3.5pt plus1pt minus1pt
\lineskiplimit=3.5pt
\parskip=4pt plus1pt minus4pt

% macro for slash
\def\negenspace{\kern-1.1em}

%\def \Behauptung{
%= \hbox to 0pt{ \kern -10pt \lower 3pt \vbox to 15pt {\hbox
%{$\scriptstyle ! \> {} $} \vglue 6pt \hbox {} }} }

%Macro for section, subsection and equation numbers:

\newcount\secno
\secno=0
\newcount\susecno
\newcount\fmno\def\z{\global\advance\fmno by 1 \the\secno.
                       \the\susecno.\the\fmno}
\def\section#1{\global\advance\secno by 1
                \susecno=0 \fmno=0
                \centerline{\bf \the\secno. #1}\par}
\def\subsection#1{\medbreak\global\advance\susecno by 1
                  \fmno=0
       \noindent{\the\secno.\the\susecno. {\it #1}}\noindent}

%Macro for d'Alembertian:

\def\sqr#1#2{{\vcenter{\hrule height.#2pt\hbox{\vrule width.#2pt
height#1pt \kern#1pt \vrule width.#2pt}\hrule height.#2pt}}}

%Macro for footnotes:

\newcount\refno
\refno=1
\def\y{\the\refno}
\def\myfoot#1{\footnote{$^{(\y)}$}{#1}
                 \advance\refno by 1}

%Macro for list of references:

%Macro for not equal:
\def\neq{\hbox{$\,$=\kern-6.5pt /$\,$}}

%Macro for = with * on top:

%Macro for boldface omega (\clom):
%AM font:
%\font\fbg=ambi10\def\clom{\hbox{\fbg\char33}}

%CM font (when using IPS):
%\font\fbg=cmmib10\def\clom{\hbox{\fbg\char33}}

%Macro for section, and equation numbers: (Use \sectio)

\newcount\secno
\secno=0
\newcount\fmno\def\z{\global\advance\fmno by 1 \the\secno.
                       \the\fmno}
\def\sectio#1{\medbreak\global\advance\secno by 1
                  \fmno=0
       \noindent{\the\secno. {\it #1}}\noindent}

%Macro for semidirect product
\def\semidirect{\;{\rlap{$\supset$}\times}\;}

%Pound sterling:= {\it \$\/}50--00.

%To use any special fonts with IPS, include the corresponding
%CM font definition in your macro or TeX file. In most cases this
%coincides with the AM font except for the first letter, e.g. cmr9
%and amr9 (note: use lower case letters). The one major exception is
%the bold symbols font, which is cmmib10 in place of ambi10, etc.

%=====================================================================
%\noindent{\it file translat.tex, 27-06-1994, Tresguerres}
%\bigskip\bigskip
\input extdef
\magnification=\magstep1
\hsize 13cm
\vsize 20cm

\hfill{Preprint IMAFF 94/2}
\centerline{\bf{NONLINEAR GAUGE REALIZATION OF SPACETIME
SYMMETRIES}}
\centerline{\bf{INCLUDING TRANSLATIONS}}
\vskip 1.0cm
\centerline{by}
\vskip 1.0cm
\centerline{J. Julve, A. L¢pez--Pinto, A. Tiemblo and R. Tresguerres}
\vskip 1.0cm
\centerline{\it {IMAFF, Consejo Superior de Investigaciones
Cient¡ficas,}}
\centerline{\it{Serrano 123, Madrid 28006, Spain}}
\vskip 1.5cm
\centerline{ABSTRACT}\bigskip
We present a general scheme for the nonlinear gauge realizations
of spacetime groups on coset spaces of the groups considered. In
order to show the relevance of the method for the rigorous
treatment of the translations in gravitational gauge theories,
we apply it in particular to the affine group. This is an illustration
of the family of spacetime symmetries having the form of a
semidirect product $H\semidirect T$, where $H$ is the stability
subgroup and $T$ are the translations . The translational
component of the connection behaves like a true tensor under $H$
when coset realizations are involved.
\bigskip\bigskip
\sectio{\bf{Introduction}}\bigskip
General Relativity has a geometrical setting. It was originally
formulated on a Riemannian space, in terms of coordinates of
spacetime and a metric tensor defined on them, determined by the
Einstein equations. Cartan's work$^{(1)}$ provided a more
suitable, coordinate--independent, framework for the geometrical
approach to theories of gravity. It is also defined on an
n--dimensional manifold given {\it{a priori}} as an
unavoidable mathematical tool, but without observable physical
meaning. There, {\it{vielbeins}} $e_\alpha:=e_\alpha {}^i
\partial _i$ are introduced to represent local frames. Parallel
transport of the reference frames is achieved by means of
1--form connections $\Gamma _\alpha {}^\beta $. From the vector
bases $e_\alpha $, coframes or bases of 1--forms $\vartheta
^\alpha := e^\alpha{}_i\, dx^i$ are defined, which are dual to
the frames with respect to the interior product $e_\alpha\rfloor
\vartheta ^\beta =\,\delta_\alpha ^\beta $. The coframes thus
play a central role in the geometrical formulation.

The geometrical interpretation of gravity is an essential
feature of the theory. Indeed, other dynamical theories, namely
the Yang--Mills theories of the remaining interactions, are
necesarily defined on the geometrical background established
by the gravitational theory. However, a unified scheme of all
interactions seems to require to treat all of them on the same
basis, namely as gauge theories. The first attempt to describe
gravity as a dynamical theory of a local spacetime group was the
work of Utiyama$^{(2)}$, where just the Lorentz group was gauged,
while the tetrads $\vartheta ^\alpha $ were introduced
{\it{a priori}} into the model as operators of reference frame
changes. The coordinates were given {\it{a priori}} too. Later
work was done to describe the tetrads as related to the gauge
potentials of some spacetime symmetry group$^{(3)(4)}$, in
particular to those of the translations. Following this spirit,
the Poincar' group was introduced as the symmetry to be gauged.
Wider groups are found in the literature$^{(5)}$, which have the
Poincar' group as a subgroup. However, several problems appear
when translations, as constitutive part of the spacetime gauge
group, are present. The physical interpretation of the
corresponding local transformations becomes dubitious, since
their gauge fields do not have any geometrical meaning. The main
difficulty is that one cannot identify them with the tetrads,
which are covectors and lack the inhomogeneous term under gauge
transformations. Some attempts were made to clarify the link
between the translational connections and the true coframes. In
the early works of Utiyama, Sciama and Kibble$^{(2)(3)}$, the
role of the translations was played by general coordinate
transformations. Hehl et al.$^{(4)(6)}$ proposed an improved
approach, based on the Poincar' group actively interpreted,
regarding the gauged translations as represented by parallel
transport. They substituted the usual translational generators
by covariant derivatives. The price one has to pay is that the
translations no longer constitute an Abelian group.

Several authors$^{(7)}$ recognized that the reformulation of
the dynamical gauge theory of gravity in terms of its standard
geometrical structure, and {\it{vice versa}}, can only be
realized by introducing extra nondynamical degrees of freedom
in the theory. In the context of Poincar' gauge theories,
Grignani and Nardelli called them the {\it{Poincar' coordinates}}
$\xi ^\alpha$. The authors considered them as Higgs--type fields
which transform as vectors under gauge transformations of the
Poincar' group. The Lorentz group is the residual gauge group
which is left invariant by the particular choice
$\xi ^\alpha =\,0$, called by the authors the {\it{physical
gauge}}, because only in this case the components of the gauge
potential along the Poincar' generators become the physical
coframe and spin connection respectively. But this restriction to
the framework of a Lorentz gauge theory of gravity is not the
only one which makes physical sense. In general as shown by
Mielke et al.$^{(8)}$, introducing a vector--valued zero form
$\xi ^\alpha$ which transforms as a Poincar' (resp. as an
affine) vector, and assuming that the relationship between the
tetrads and the linear translational gauge fields
${\buildrel (T)\over{\Gamma ^\alpha}}$ is given by
$$\vartheta ^\alpha :=D\xi ^\alpha
+{\buildrel (T)\over{\Gamma ^\alpha}}\,,\eqno(\z)$$
the right transformation properties are guaranteed. Hayashi et
al.$^{(9)}$ considered earlier a simplified version of this
definition of the tetrads.

One of the goals of the present paper is to explain the origin
of the {\it{coordinates}} $\xi ^\alpha $ mentioned above as
coset parameters which, as we will show below, come out from the
nonlinear approach to the gauge theory of spacetime groups.
Indeed, in the global limit, they become indistinguishable from
Cartesian coordinates {\it{via}} the identification
$\xi ^\alpha =\,\delta ^\alpha _i x^i$. They turn out to be an
essential element in the definition of the nonlinear connection.
We will show that the term $D\xi ^\alpha $ added to the
translational connection to construct the coframe, see (1.1), is
a necessary contribution arising from the nonlinear realization
of spacetime gauge symmetries in general.

We propose a scheme in which the coframes turn out to be
nonlinear connections associated to the generators of the
coset, with the right tensorial transformation properties.
The translational connection does not appear as an
independent object of the theory, since the translations
are realized nonlineary, so that the translational
covariance is present but does not become explicitely
apparent. Only the particular combination (1.1) of the
translational gauge fields and the coset parameters occurs,
playing the role of the coframes. The whole geometrical setting
arises as the result of gauging the nonlinear realization of the
group in a coset space, without making recourse to any
{\it{a priori}} structure other than the basic symmetry group.
Indeed, nonlinear group realizations allow to interpret a
quotient space of the whole group space as the very spacetime
manifold, as noticed by several authors$^{(10)}$. The coset
parameters are interpreted as spacetime coordinates. The
coordinate--independent formulation of the theory in terms of
differential forms thus shows that the theory is independent of
the coset parametrization. In our approach, the translational
generators obey the usual commutation relations.

\bigskip
\sectio{\bf{Nonlinear realization of spacetime groups}}\bigskip

The coset representation introduced by Coleman et al.$^{(11)}$
in the context of phenomenological Lagrangians is based on
the non linear action of a group on itself. More precisely, the
group acts on the cosets defined with respect to a subgroup $H$
which classifies the fields of the theory. The coset parameters
play the role of the spacetime manifold. The explicit gauge
covariance with respect to the classification subgroup is
maintained. The technique, initially proposed to treat
internal symmetries, was soon extended to spacetime
symmetries. In fact, several attempts were made to apply the
nonlinear realizations to gravity. Isham, Salam and
Strathdee$^{(12)}$ considered the nonlinear action of
$GL(4\,,R)$, taking the Lorentz group as classification
subgroup, see below, and Borisov and Ogievetski$^{(13)}$
proposed to require covariance under simultaneous nonlinear
realizations of the affine and the conformal groups.
But none of these works solved the main problem
of explaining the relationship between the tetrads and the
translational gauge fields since they only considered the global
group action. Instead, a symmetric tensor constructed from
the parameters of the symmetric affine transformations was
identified as the "vierbein field". Stelle and West$^{(14)}$
investigated the nonlinear realization allowed
by the spontaneous symmetry breaking of SO(3,2) down to SO(3,1).
Pseudotranslations were defined from the broken generators of
SO(3,2). Their parameters, i.e. the nondynamical SO(3,2) vector
fields $\xi ^\alpha $ constrained to take their values in an internal
anti--de Sitter space, were identified as the Goldstone fields
associated to the symmetry breaking, not as coordinates. Chang
and Mansouri$^{(15)}$ made use of the general nonlinear approach
as we do, but they did not emphasize its relevance for the gauge
teatment of the translations and thus for the link between
tetrads and translational gauge fields. They also introduced an
auxiliary coordinate manifold alien to the group. Volkov and
Soroka$^{(10)}$ considered a nonlinear
realization of the Poincar' group in the context of the
spontaneous breakdown of supersymmetry. They identified the
coset parameters with the points of spacetime itself.
Lord$^{(10)}$ interpreted the gauge generalization of a
spacetime group $G$ in the language of fiber bundles as a
gauge generalization of the stability subgroup, together with
diffeomorphisms, following Hehl et al.$^{(5)}$. The translations
thus loose their Abelian character. According to him, the
gauge potentials on the coset space are the pullbacks of the
connection on $G$, while we define them as the
generalized Maurer--Cartan 1--form given in eq.(3.4).

Our treatment resembles more closely that of Chang and
Mansouri$^{(15)}$, but we identify the coset space with spacetime
itself, as Lord and others$^{(10)}$ do. We claim that the
coset parameters do not play the role of any kind of field,
but they must be considered as the coordinates themselves,
as we have mentioned before.

Here we will outline the basic nonlinear machinery briefly.
Let $G=\{g\}$ be a connected, semisimple Lie group, and
$H=\{h\}$ a subgroup of $G$. We assume that linear
representations $\rho (h)$ of the classification subgroup
$H$ exist, acting on functions $\psi $ belonging to a
representation space of $H$. One defines the action of $G$
on the coset space $G/H$. The "points" of $G/H$ are equivalence
classes of the form $cH$, with $c\epsilon\left(G-H\right)$.
Because we are dealing with Lie groups, the elements of $G/H$ are
labeled by continuous parameters, say $\xi$. We choose the coset
indicators $c(\xi )$ parametrized by $\xi$ as the
representatives of the points of
$G/H$. As we will see, these coset parameters will play the role
of a kind of coordinates. Now we let act the group elements
$g\epsilon G$ on $G/H$ according to
$$\eqalign{g:\, G/H&\rightarrow G/H\cr
c\,(\xi )&\rightarrow c\,(\xi ')\,,\cr }\eqno(\z)$$
according to the general law
$$g\,c\,(\xi\,)=\, c\,(\xi ')\, h\left(\xi\,, g\right)\,.\eqno(\z)$$
Moreover, eq.(2.2) defines the group element $h\left(\xi\,,
g\right) \epsilon H$
governing the behavior of the fields $\psi$ under $G$.
The elements $g$ of the whole group $G$ considered in (2.2)
act nonlinearly on the representation space of the
classification subgroup $H$ according to
$$\psi '=\,\rho\left(h\left(\xi\,, g\right)\right)\psi\,.\eqno(\z)$$
Formally, eq.(2.3) resembles the linear action of $H$, but the
nonlinearity manifests itself in general through $h\left(\xi\,,
g\right)$ as given by eq.(2.2).
Generally $h$ and therefore $\psi $ itself are
functions of $\xi $. The action of the group is realized on the pairs
$\left(\xi\,,\psi\right)$. It reduces to the usual
linear action of $H$ when we take in particular for $g$ in (2.2) an
element of $H$.

\bigskip
\sectio{\bf{Local theory}}\bigskip

Now we will discuss the main point of this paper. Our aim is to
construct the connection suitable to define a covariant
differential transforming like (2.3) under local
transformations. In other words, we look for the connection
associated to the nonlinear realization of G, which will behave
like a connection of the classification subgroup $H$. Even in the
global case, in which $g$ does not depend on $\xi$,
the appearence of $\xi$ in the transformation $h\left(\xi\,,
g\right)$, see (2.2), implies that $h$ is not a constant, and a
particular kind of covariant differentials is required, namely
$${\buildrel o\over {D}}\psi :=\left( d\, +{\buildrel
o\over{\Gamma }} \,\right)\psi\,,\eqno(\z)$$
where ${\buildrel o\over{\Gamma }}$ is the Maurer--Cartan
connection 1--form with values on the group algebra
$${\buildrel o\over{\Gamma }}:=\, c^{-1} d\, c\,,\eqno(\z)$$
transforming as
$${\buildrel o\over{\Gamma }}'=\,h
{\buildrel o\over{\Gamma }} h^{-1} +h d\, h^{-1}\,.\eqno(\z)$$
We emphasize that, when local transformatios are involved, the
previous scheme must be extended to include a dependence of
the parameters of $g\epsilon G$ on the {\it{coset coordinates}}
$\xi$. We then generalize (3.2) to the connection corresponding
to the nonlinear realization of the group, namely
$$\Gamma:=\, c^{-1}{\rm {\cal{D}}}c\,,\eqno(\z)$$
where the covariant differential on the coset space is defined as
$${\rm {\cal{D}}}c:=\,\left(d\,+\Omega\,\right)c\,,\eqno(\z)$$
with the ordinary linear connection $\Omega$ of the whole group G
transforming as
$$\Omega '=\, g\,\Omega \,g^{-1}+g\,d\,g^{-1}\,.\eqno(\z)$$
It is easy to prove that the nonlinear gauge field $\Gamma$
defined in (3.4) transforms as
$$\Gamma '=\,h \Gamma h^{-1} +h d\, h^{-1}\,,\eqno(\z)$$
thus allowing to write true local covariant differentials of the
$\psi$ fields of the theory as
$${\bf D}\psi :=\left( d\,
+\Gamma\,\right)\psi \,,\eqno(\z)$$
obeying the same transformation law (2.3) under the local
action of $g\epsilon G$. The components of the connection
$\Gamma$ have very intersting transformation properties. In
fact, it is easy to read out from (3.7) that

\noindent {\bf a.-} only the components of $\Gamma$ related to
the generators of $H$ behave as true connections, i.e. transform
inhomogeneously,

\noindent {\bf b.-} the components of $\Gamma$ over the
generators associated with the cosets $c$ transform as tensors
with respect to the subgroup $H$ notwithstanding their nature of
connections.

We notice that these properties provide us with a general scheme
which reproduces the main features of the gauge versions of gravity.
In it the translational components of the connection appear
as pure tensors, as we are going to see in the next section.

\bigskip
\sectio{\bf{Nonlinear gauge approach to the affine group}}\bigskip

As a relevant example, let us consider the affine group
$A(n\,,R)=\, GL(n\,,R)\semidirect R^{n}$ in $n$ dimensions,
defined as the semidirect product of the translations and the
general linear transformations. The following discussion is also
appliable to the Poincar' group. The commutation relations loud
$$\eqalign{\left[L^\alpha {}_\beta\,,L^\mu {}_\nu\right]=&
\,i\,\left(\delta^\alpha _\nu L^\mu {}_\beta
         -\delta^\mu _\beta L^\alpha {}_\nu\right)\,,\cr
\left[L^\alpha {}_\beta\,, P_\mu\right]\hskip0.15cm=&
\,i\,\delta^\alpha _\mu P_\beta\,,\cr
\left[P_\alpha\,, P_\beta\right]\hskip0.25cm
=&\,0\,,\cr}\eqno(\z)$$
where $P_\alpha $ are the generators of the
translations, and $L^\alpha {}_\beta $ those of the linear
transformations. We will realize the group action on the coset space
$A(n\,,R)/GL(n\,,R)$. We choose in particular for the cosets the
following parametrization:
$$c:=e^{-i\,\xi ^\alpha P_\alpha}\,,\eqno(\z)$$
where $\xi ^\alpha$ are the {\it{coset coordinates}}. As we will
see below, they are equivalent to {\it{Cartan's generalized
radius vector}} or the {\it{Poincar' coordinates}} considered by
other authors$^{(7)(8)}$. The group elements of the whole affine
group $A(n\,,R)$ are parametrized as
$$g=\,  e^{i\,a^{\alpha} P_\alpha}
e^{i\,u_\alpha {}^\beta L^\alpha {}_\beta}\,,\eqno(\z)$$
and those of the classification subgroup $GL(n\,,R)$ are taken to
be
$$h:=e^{i\,u '_\alpha {}^\beta L^\alpha {}_\beta}\,.\eqno(\z)$$
Other parametrizations which lead to equivalent results are of
course possible. The fundamental eq.() defining the nonlinear
group action then reads
$$e^{i\,a^{\alpha} P_\alpha}
e^{i\,u_\alpha {}^\beta L^\alpha {}_\beta}
e^{-i\,\xi ^\alpha P_\alpha} =\, e^{-i\,\xi ^{'\alpha} P_\alpha}
e^{i\,u '_\alpha {}^\beta L^\alpha {}_\beta}\,.\eqno(\z)$$
Repeatedly using the Hausdorff's formula, the explicit
expressions for the transformed coset parameter
$\xi ^{'\alpha}$ and for $u '_\alpha {}^\beta $, in the r.h.s. of
(4.5), are calculable. After a little algebra, we get
$$\xi ^{'\alpha }=\,\left(\Lambda ^{-1}\right) _\beta {}^\alpha
\,\xi ^\beta -a^\alpha\quad\,,\qquad u '_\alpha {}^\beta
=\, u_\alpha {}^\beta \,.\eqno(\z)$$
In eq.(4.6) and in the following, we use the definitions
$$\Lambda  _\beta {}^\alpha :=\, e^{u_\beta {}^\alpha }\quad
\,,\qquad\left(\Lambda ^{-1}\right) _\beta {}^\alpha
:=\, e^{-u_\beta {}^\alpha }\,.\eqno(\z)$$
These matrices which stand for the regular representation of (4.4)
describe the two possible actions of the group associated to
covariant and contravariant characters. Thus we see from (4.6)
that the coset parameters $\xi ^\alpha$ transform as affine
covectors, as postulated by other authors$^{()}$ for {\it{Cartan's
generalized radius vector}}. We notice that in the global case
the differentials of the coordinates transform as contravariant
$GL(n\,,R)$ vectors, as seen from (4.6), so that the
distinction between covariant and contravariant tensors is
already present in the scheme. Observe that for the particular
choice of the coset space we are dealing with, the parameters
$u '_\alpha {}^\beta$ of the r.h.s. of (4.5), i.e. of the
parameters characterizing $h\epsilon H$, according to the
general formulation (2.2), coincide with those $u_\alpha {}^\beta$
of the parametrization of $g\epsilon G$ in the l.h.s. of (4.5).
Although this result is not valid for arbitrary choices of the
coset space, it simplifies things in our case since
the action (2.3) of the whole affine group on arbitrary fields of a
representation space of $GL(n\,,R)$ reduces to
$\psi '=\,\rho\left(h\left( g\right)\right)\psi\,.$
Actually we have
$$\psi '=\,\rho\left(\Lambda\right)\psi\,,\eqno(\z)$$
with
$$\rho\left(\Lambda\right):=\,e^{i\, u_\alpha {}^\beta\rho
\left(L^\alpha{}_\beta\right)}\eqno(\z)$$
being an arbitrary representation of $GL(n\,,R)$ transformations.
Now we define the suitable connection for the
nonlinear gauge realization in two steps, first introducing the
ordinary linear affine connection $\Omega $ in (3.5) as
$$\Omega :=-i\,{\buildrel (T)\over{\Gamma ^\alpha}} P_\alpha
         -i\,\Gamma _\alpha {}^\beta L^\alpha {}_\beta\,,\eqno(\z)$$
which includes the true translational potential
${\buildrel (T)\over{\Gamma ^\alpha}}$ and the $GL(n\,,R)$
connection $\Gamma _\alpha {}^\beta $.
The transformations (3.7) take the standard form
$$\Gamma '_\alpha {}^\beta =\,\left(\Lambda ^{-1}\right) _\gamma {}^\beta
\Gamma _\delta {}^\gamma \Lambda  _\alpha {}^\delta
+\left(\Lambda ^{-1}\right) _\gamma {}^\beta d\,\Lambda  _\alpha {}^\gamma
\,,\eqno(\z)$$
and
$${\buildrel (T)\over{\Gamma ^{'\alpha}}}
=\,\left(\Lambda ^{-1}\right) _\beta {}^\alpha\,
\left[\,{\buildrel (T)\over{\Gamma ^\beta}}\, +D\left(
\Lambda _\gamma {}^\beta a^\gamma\right)\,\right]\,,\eqno(\z)$$
with $D$ as the covariant differential constructed with the
$GL(n\,,R)$ connection exclusively. Making then use of
definition (3.4), we get
$$\Gamma :=\,e^{i\,\xi ^\alpha P_\alpha}\left(d\,+\Omega
\,\right) e^{-i\,\xi ^\alpha P_\alpha} =-i\,\left(D\xi ^\alpha
+{\buildrel (T)\over{\Gamma ^\alpha}}\right) P_\alpha
-i\,\Gamma _\alpha {}^\beta L^\alpha {}_\beta\,.\eqno(\z)$$
The components on $P_\alpha $ in the connection (4.13) play a
crucial role in gravitational gauge theories. As we mentioned
before, in spite of the fact that they arise as a constitutive
part of the connection required in the nonlinear realization,
they do not transform as a connection, but as a covector of the
classification subgroup. Let us define them as the coframe
$$\vartheta ^\alpha :=D\xi ^\alpha
+{\buildrel (T)\over{\Gamma ^\alpha}}\,.\eqno(\z)$$
According to (3.7), it transforms as a covector under
$GL(n\,,R)$. Applying (4.6,11,12), we find explicitely
$$\vartheta ^{'\alpha }=\,\left(\Lambda ^{-1}\right) _\beta
{}^\alpha \vartheta ^\beta\,.\eqno(\z)$$
The coframe (4.14) provides the link between the dynamical approach
and the geometrical interpretation of the formalism, since it
plays the role of the basis of 1--forms. In terms of
$\vartheta ^\alpha $, we geometrize the dynamical theory
defining the vector basis $e_\alpha$ by means of the general relation
$$e_\alpha\rfloor\vartheta ^\beta =\,\delta _\alpha ^\beta\,.\eqno(\z)$$
The vector basis transforms as a vector, namely
$$e'_\alpha =\,\Lambda  _\alpha {}^\beta e_\beta\,.\eqno(\z)$$
The general linear connection in (4.13) transforms as before,
see (4.11).

\noindent Commutation of two covariant differentials (3.8) yields
$${\bf D}\wedge {\bf D} =-i\, T^\alpha P_\alpha
-i\,R _\alpha {}^\beta L^\alpha {}_\beta\,,\eqno(\z)$$
with the torsion $T^\alpha $ and the curvature $R _\alpha
{}^\beta $ respectively defined as
$$T^\alpha :=\,D\vartheta ^\alpha \,,\eqno(\z)$$
and
$$R _\alpha {}^\beta :=\, d\Gamma _\alpha {}^\beta
+\Gamma _\gamma {}^\beta\wedge\Gamma _\alpha {}^\gamma \,,\eqno(\z)$$
thus showing the groupal character of the torsion as the field
strenght of the translations.\bigskip

\sectio{\bf{Final remarks}}\bigskip

We notice that in our scheme the expression (4.14) reproduces
the same results obtained in reference (8) without any
{\it{ad hoc}} assumptions. We claim that the general scheme of gauge
nonlinear coset realizations provides the natural framework for
the gauge theories of gravity. The restriction to the Poincar'
group follows exactly the same lines as the treatment of the
affine group and consequently we do not need to insist on.
\bigskip
\centerline {\bf Acknowledgement}\bigskip
\noindent We are very grateful to Prof. F.W. Hehl for helpful comments.
\vskip1.0cm
\centerline{REFERENCES}\vskip1.0cm

\noindent [1] E. Cartan, {\it Sur les vari't's … connexion
affine et la th'orie de la relativit' g'n'ralis'e}, Ouvres
completes, Editions du C.N.R.S. (1984), Partie III. 1, pgs. 659
and 921

\noindent [2] R. Utiyama,  {\it Phys. Rev.} {\bf 101} (1956) 1597

\noindent [3] T. W. B. Kibble, {\it J. Math. Phys.} {\bf 2} (1961) 212

D. W. Sciama, {\it Rev. Mod. Phys.} {\bf 36} (1964) 463 and 1103

\noindent [4] F. W. Hehl, P. von der Heyde, G. D. Kerlick and J. M. Nester,
{\it Rev. Mod. Phys.} {\bf 48} (1976) 393

\noindent [5] A. G. Agnese and P. Calvini, {\it Phys. Rev.} {\bf
D 12} (1975) 3800 and 3804

F. W. Hehl, J. D. McCrea,  E. W. Mielke, and Y. Ne'eman
{\it Found. Phys.} {\bf 19} (1989) 1075

R. D. Hecht and F. W. Hehl,  {\it {Proc. 9th Italian Conf.
G.R. and Grav. Phys. , Capri (Napoli) }}. R. Cianci
et al.(eds.) (World Scientific, Singapore, 1991) p. 246

F.W. Hehl, J.D. McCrea, E.W. Mielke, and Y. Ne'eman, {\it
Physics Reports}, to be published.

\noindent [6] P. von der Heyde, {\it Phys. Lett.} {\bf 58 A} (1976) 141

\noindent [7] G. Grignani and G. Nardelli, {\it Phys. Rev.}
{\bf D 45} (1992) 2719

H.R. Pagels, {\it{Phys. Rev.}} {\bf D 29} (1984) 1690

T. Kawai, {\it{Gen. Rel. Grav.}} {\bf 18} (1986) 995

(?) A. Trautman, in {\it Differential Geometry},
Symposia Mathematica Vol. 12 (Academic Press, London, 1973), p. 139

\noindent [8] E. W. Mielke, J.D. McCrea, Y. Ne'eman and F.W. Hehl
{\it Phys. Rev.} {\bf D 48} (1993) 673, and references therein

\noindent [9] K. Hayashi and T. Nakano, {\it Prog. Theor. Phys}
{\bf 38} (1967) 491

K. Hayashi and T. Shirafuji, {\it Prog. Theor. Phys}
{\bf 64} (1980) 866 and {\bf 80} (1988) 711

\noindent [10] D.V. Volkov and V.A. Soroka, {\it JETP Lett.} {\bf 18}
(1973) 312, and {\it Theor. Math. Phys.} {\bf 20} (1975) 829

E.A. Lord, {\it{Gen. Rel. Grav.}} {\bf 19} (1987) 983

\noindent [11] S. Coleman, J. Wess and B. Zumino, {\it Phys. Rev.}
{\bf 117} (1969) 2239

C.G. Callan, S. Coleman, J. Wess and B. Zumino, {\it Phys. Rev.}
{\bf 117} (1969) 2247

\noindent [12] A. Salam and J. Strathdee, {\it Phys. Rev.}
{\bf 184} (1969) 1750 and 1760

C.J. Isham, A. Salam and J. Strathdee, {\it Ann. of Phys.}
{\bf 62} (1971) 98

\noindent [13] A.B. Borisov and V.I. Ogievetskii, {\it Theor.
Mat. Fiz.} {\bf 21} (1974) 329

\noindent [14] K.S. Stelle and P.C. West, {\it Phys. Rev.}
{\bf D 21} (1980) 1466

\noindent [15] L.N. Chang and F. Mansouri, {\it Phys. Lett.} {\bf 78 B} (1979)
274; {\it Phys. Rev.} {\bf D 17} (1978) 3168

\noindent [] G.A. Sardanashvily  and M. Gogbershvily, {\it Mod.
Phys. Lett.} {\bf A 2} (1987) 609

\noindent [] G.A. Sardanashvily, Preprint gr-qc/9405013
\vfill\eject

\bye